\documentclass[%
 reprint,
superscriptaddress,
showpacs,
preprintnumbers,
 amsmath,amssymb,
pra,
longbibliography
]{revtex4-1}

\usepackage{graphicx}
\usepackage{dcolumn}
\usepackage{bm}
\usepackage{setspace}	
\usepackage{color}
\usepackage{amsmath}	

\DeclareGraphicsExtensions{.eps}
\graphicspath{{figs/}}

\def\ii{{\mathrm{i}}}

\def\bra#1{\langle #1|}

\def\ket#1{|#1\rangle}

\def\bracket#1{\langle #1 \rangle}
\def\bracketi#1#2{\langle #1 | #2 \rangle}
\def\bracketii#1#2#3{\langle #1 | #2| #3\rangle}

\def\sub#1{_\mathrm{#1}} 
\def\sur#1{^\mathrm{#1}} 

\def\tr{\mathrm{tr}}

\begin{document}


\title{
A unified view of direct measurement of quantum states, processes, and measurements
}


\author{Kazuhisa Ogawa}
\email{k-ogawa.qiqb@osaka-u.ac.jp}
\affiliation{%
Center for Quantum Information and Quantum Biology (QIQB), Osaka University, Osaka 560-0043, Japan
}%
 
\author{Takumi Matsuura}
\affiliation{%
Graduate School of Information Science and Technology, Hokkaido University, Sapporo 060-0814, Japan
}%

\author{Akihisa Tomita}
\affiliation{%
Graduate School of Information Science and Technology, Hokkaido University, Sapporo 060-0814, Japan
}%

\date{\today}

\begin{abstract}
The dynamics of a quantum system are characterized by three components: quantum state, quantum process, and quantum measurement.
The proper measurement of these components is a crucial issue in quantum information processing.
Recently, direct measurement methods have been proposed and demonstrated wherein each complex matrix element of these three components is obtained separately, without the need for quantum tomography of the entire matrix. 
Since these direct measurement methods have been proposed independently, no theoretical framework has been presented to unify them despite the time symmetry of quantum dynamics.  
In this study, we propose a theoretical framework to systematically derive direct measurement methods for these three components. 
Following this framework and further utilizing the basis-shift unitary transformation, we have derived the most efficient direct measurement method using qubit probes. 
Additionally, we have experimentally demonstrated the feasibility of the direct measurement method of quantum states using optical pulse trains.
\end{abstract}

\maketitle

\begin{spacing}{1.0}

\section{Introduction}


The dynamics of a quantum system are characterized by three key components: the quantum state, the quantum process, and the quantum measurement. 
Each of these components is represented by specific quantum elements: the density operator $\hat{\rho}$, the completely positive and trace-preserving (CPTP) map $\mathcal{M}$, and the positive-operator-valued measure (POVM) element $\hat{E}$, respectively.  
These components are typically represented by complex matrices on a fixed basis, and the appropriate measurement of these matrix elements is a fundamental issue in various quantum applications. 
The standard method for measuring quantum states is known as quantum state tomography (QST) \cite{PhysRevA.40.2847,PhysRevLett.70.1244,breitenbach1997measurement,PhysRevLett.83.3103,PhysRevA.64.052312}.
In QST, projective measurements are performed on different bases, and each matrix element of the density operator $\hat{\rho}$ is estimated by post-processing all the measured data. 
Similarly, for quantum processes and quantum measurements, methods such as quantum process tomography (QPT)
\cite{PhysRevLett.78.390,chuang1997prescription,PhysRevA.64.012314,PhysRevLett.91.120402} and quantum measurement tomography (QMT) \cite{PhysRevLett.106.020502,PhysRevA.98.042318,PhysRevResearch.5.033154} are utilized. Analogous to QST, the CPTP map $\mathcal{M}$ and the POVM element $\hat{E}$ can be determined by preparing various initial states, conducting projection measurements on various bases, and subsequently post-processing all the measurement data. 
In quantum tomography, as the dimension of the Hilbert space of the measured quantum system increases, the number of measurements and the complexity of the reconstruction algorithm also increase.
This may render the quantum tomography impractical for certain applications.
However, for some quantum applications \cite{RevModPhys.81.865,PhysRevLett.125.060404,PhysRevLett.120.170502,PhysRevLett.106.230501,PhysRevLett.107.210404,huang2020predicting}, it is not imperative to reconstruct the complete matrix; rather, only specific parts of it need to be measured directly. 
In such scenarios, it becomes crucial to develop methods for directly measuring only certain matrix elements of the quantum component under examination. 


Recently, a new class of measurement methods, referred to as direct measurements (DM), has been emerged, exhibiting distinct characteristics from various quantum tomography methods. 
In DM, each matrix element of the quantum component under investigation can be estimated within a single measurement setup, eliminating the need for post-processing the results of measurements across different initial states or measurement bases. 
Initially, DM was applied to pure-state wavefunctions \cite{lundeen2011direct} using the Aharonov--Albert--Vaidman (AAV) weak measurement \cite{PhysRevLett.60.1351}. 
Subsequent studies expanded the scope of DM to encompass objects such as pseudo-probability distributions \cite{PhysRevLett.112.070405}, density operators \cite{PhysRevLett.117.120401}, nonlocal entangled states \cite{PhysRevLett.123.150402}, quantum processes \cite{kim2018direct,gaikwad2023direct}, and quantum measurements \cite{PhysRevLett.127.180401}. 
Notably, certain DM techniques have been implemented not through AAV weak measurements, but via standard indirect measurements employing strong interactions and projective measurements on the probe system, resulting in higher measurement efficiency, as experimentally demonstrated \cite{PhysRevLett.116.040502,PhysRevLett.121.230501,ogawa2021direct,PhysRevLett.127.040402,PhysRevA.104.042403,PhysRevA.105.062414,PhysRevA.101.012119,PhysRevLett.127.030402,PhysRevLett.123.190402,xu2021direct}.


In prior studies, DM methods have been independently proposed for each of the three quantum components. 
However, given the time-reversal symmetry of quantum time evolution, it is reasonable to consider direct measurement methods for these components from a more unified perspective.
Previously, we introduced an intuitive diagrammatic representation of DM for wavefunctions using qubit probes, elucidating how DM operates regardless of measurement strength \cite{ogawa2019framework}. 
In this research, we expand upon this framework and demonstrate the systematic design of efficient DM systems for quantum states, quantum processes, and quantum measurements. 
This framework can be viewed as a generalized Hadamard test and offers guidance for designing quantum circuits to obtain desired complex observables. 
Leveraging this framework, we introduce a novel DM method employing the basis-shift unitary transformation, which proves to be more efficient per shot measurement compared to previous DM methods. 
Moreover, we validate the feasibility of this proposed DM method for quantum states (density operators) by experimentally applying it to optical pulse trains encoded in discrete time degrees of freedom. 


This paper is structured as follows: 
In Section~\ref{sec:2}, we introduce the generalized Hadamard test along with its diagrammatic representation, illustrating its applicability in deriving DM methods for various quantum components. 
In Section~\ref{sec:3}, we detail the experimental demonstration of applying the newly derived direct measurement method for density operators to optical pulse trains. 
Finally, in Section~\ref{sec:4}, we provide a discussion and summary of this study. 



\section{Theory}\label{sec:2}

\subsection{Generalized Hadamard test and its diagrammatic representation}

\begin{figure}[t]		
\centering
\includegraphics[width=8.5cm]{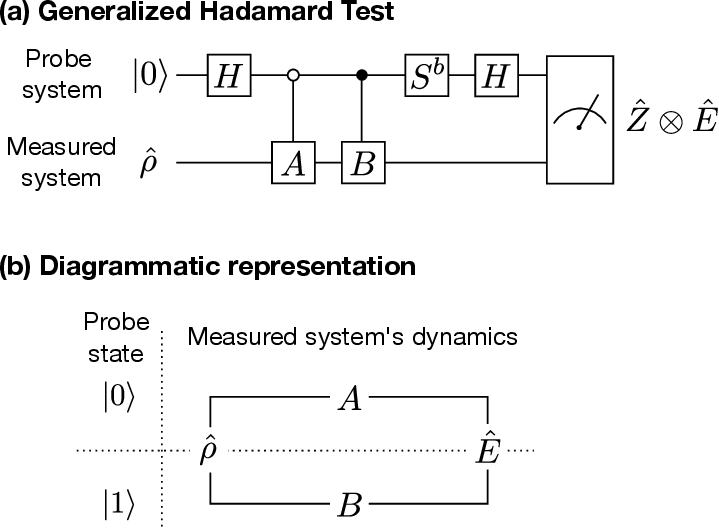}
\caption{
(a) Quantum circuit illustrating the generalized Hadamard test. 
The expectation values of $\hat{Z}\otimes\hat{E}$ for $b=0,1$ correspond to the real and imaginary parts of $\tr(\hat{A}\hat{\rho}\hat{B}^\dag\hat{E})$ respectively. 
(b) Diagrammatic representation for intuitive comprehension of the generalized Hadamard test. 
The top and bottom rows of the diagram depict the time evolution of the measured system when the probe system is $\ket{0}$ and $\ket{1}$, respectively. 
The initial states of the probe system are set to $\ket{+}$, and finally $\hat{X}$- and $\hat{Y}$-basis measurements are conducted; however, these are not explicitly depicted in the diagrammatic representation.
The complex value $\tr(\hat{A}\hat{\rho}\hat{B}^\dag\hat{E})$ obtained from the measurement can be intuitively understood as the trace of the operator $\hat{A}\hat{\rho}\hat{B}^\dag\hat{E}$, forming a cyclic product of the operators $\hat{\rho}$, $\hat{A}$, $\hat{E}$, and $\hat{B}$ in the diagram.
}
\label{fig:1}
\end{figure}

To systematically derive DM methods for complex matrix elements of quantum components, we first explore the generalized Hadamard test. 
This test is implemented through a quantum circuit comprising a qubit probe system and a measured system, as depicted in Fig.~\ref{fig:1}(a). 
The initial states of the probe system and the measured system are $\ket{0}\bra{0}$ and $\hat{\rho}$, respectively. 
The quantum gates include the first Hadamard gate $\hat{H}$, two controlled gates ($\ket{0}$-controlled $\hat{A}$ and $\ket{1}$-controlled $\hat{B}$), the phase gate $\hat{S}^b$ (an identity gate for $b=0$ and a phase gate for $b=1$), and a second Hadamard gate $\hat{H}$. 
Following the passage through these quantum gates, we measure the expectation value of the tensor product $\hat{Z}\otimes\hat{E}$, composed of the Pauli-Z operator $\hat{Z}$ of the probe system and a POVM element $\hat{E}$ of the measured system. 
Depending on whether $b$ is 0 or 1, the resulting expectation value is obtained by this measurement is given by 
\begin{align}
\bracket{\hat{Z}\otimes\hat{E}} 
&= 
\left\{
\begin{array}{ll}
\mathrm{Re}\left[\tr(\hat{A}\hat{\rho}\hat{B}^\dag\hat{E})\right]
& (b=0),\vspace{1mm}\\
\mathrm{Im}\left[\tr(\hat{A}\hat{\rho}\hat{B}^\dag\hat{E})\right]
& (b=1).
\end{array}
\right.
\end{align} 
In particular, the case $\hat{A}=\hat{E}=\hat{I}$ corresponds to the usual Hadamard test. 
By combining these two results, the complex value $\tr(\hat{A}\hat{\rho}\hat{B}^\dag\hat{E})$ is obtained by the generalized Hadamard test. 


The value $\tr(\hat{A}\hat{\rho}\hat{B}^\dag\hat{E})$ obtained through the generalized Hadamard test can be intuitively grasped using the diagrammatic representation introduced in Ref.~\cite{ogawa2019framework}, as depicted in Fig.~\ref{fig:1}(b). 
In this representation, certain elements such as the initial state $\ket{0}$, quantum gates $\hat{H}$ and $\hat{S}^b$, and the quantum measurement $\hat{Z}$ in the probe system are not explicitly depicted as they are assumed, while only the initial state of the measured system, the control gates dependent on the probe state, and the POVM elements of the measurement are displayed on the diagram. 
The complex value $\tr(\hat{A}\hat{\rho}\hat{B}^\dag\hat{E})$ acquired via the generalized Hadamard test can be interpreted as the trace of the operator $\hat{A}\hat{\rho}\hat{B}^\dag\hat{E}$, which forms a cyclic product of the operators $\hat{\rho}$, $\hat{A}$, $\hat{E}$, and $\hat{B}$ in the diagram (noting that the operator $\hat{B}$ in the bottom row of the diagram must take the Hermitian conjugate). 
Leveraging this relationship, we can derive a DM method for a desired complex value by appropriately arranging the operators on the diagram.

\subsection{Derivation of direct measurement methods for quantum components}

\begin{figure}[t]		
\centering
\includegraphics[width=9cm]{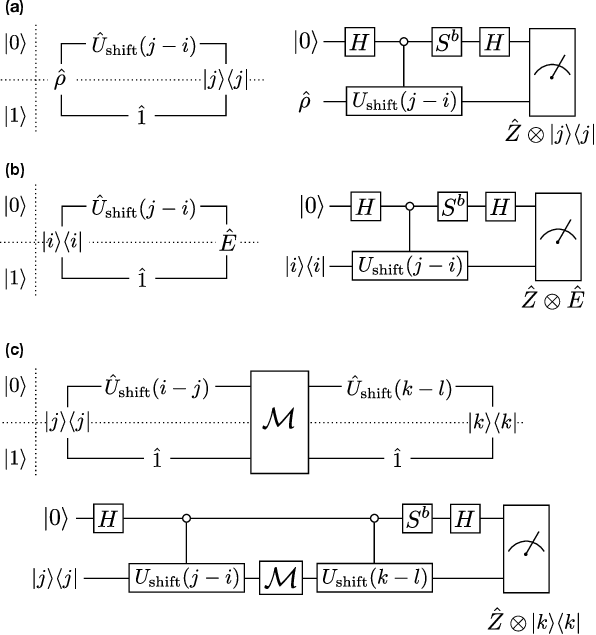}
\caption{
(a) Diagrammatic representation illustrating the quantum dynamics enabling the DM of the matrix elements of the quantum state $\hat{\rho}$, along with the corresponding quantum circuit. 
(b) Diagrammatic representation demonstrating the DM of the matrix elements of the measurement POVM element $\hat{E}$, accompanied by the corresponding quantum circuit. 
(c) Diagrammatic representation showcasing the DM of the matrix elements of a quantum process $\mathcal{M}$, along with the corresponding quantum circuit.
}
\label{fig:2}
\end{figure}

Using the diagrammatic representation, we derive a generalized-Hadamard-test-based DM method for each quantum component---quantum states, measurement POVM elements, and quantum processes. 
First, to directly measure the matrix element $\rho_{ij}:=\bracketii{i}{\hat{\rho}}{j}=\tr(\hat{\rho}\ket{j}\bra{i})$ of the quantum state $\hat{\rho}$, the diagrammatic representation suggests that we can select $\hat{A}$, $\hat{E}$, and $\hat{B}$ such that $\hat{B}^\dag\hat{E}\hat{A}\propto\ket{j}\bra{i}$ holds. 
For example, as in Refs.~\cite{PhysRevLett.117.120401,PhysRevLett.121.230501}, if we choose $\hat{A} = \ket{i}\bra{i}$, $\hat{B} = \ket{j}\bra{j}$, and $\hat{E}=\ket{\mathrm{MUB}}\bra{\mathrm{MUB}}$, the above condition is satisfied, yielding $\tr(\hat{A}\hat{\rho}\hat{B}^\dag\hat{E}) = \rho_{ij}/d$, where $\ket{\mathrm{MUB}}$ represents the mutually unbiased basis of $\ket{i}$ and $\ket{j}$, and $\bracketi{i}{\mathrm{MUB}}=\bracketi{j}{\mathrm{MUB}}=1/\sqrt{d}$ (where $d$ is the dimension of the measured system). 
As another example, to maximize the coefficient of $\ket{j}\bra{i}$ and enhance measurement efficiency, it is preferable to choose $\hat{A}$ and $\hat{B}$ as unitary operators. 
For instance, selecting $\hat{A}=\hat{U}\sub{shift}(j-i)=\sum_k\ket{k+j-i\ \mathrm{mod}\ d}\bra{k}$, $\hat{B}=\hat{1}$, and $\hat{E}=\ket{j}\bra{j}$, as illustrated in Fig.~\ref{fig:2}(a), eliminates the coefficient of $1/d$, resulting in $\tr(\hat{A}\hat{\rho}\hat{B}^\dag\hat{E}) = \rho_{ij}$. 
Here, $\hat{U}\sub{shift}(n)$ represents the basis-shift unitary transformation defined by 
\begin{align}
\hat{U}\sub{shift}(n):=\sum_k\ket{k+n\ \mathrm{mod}\ d}\bra{k},
\end{align}
which shifts the basis number by $+n$. 
Hence, the generalized Hadamard test can be utilized to determine the expectation value of a non-physical observable that does not meet the definition of a measurement POVM element, such as $\ket{j}\bra{i}$ for the state $\hat{\rho}$. 
In the latter part of this paper, we will conduct experiments to demonstrate the DM method of the quantum state using this basis-shift unitary transformation $\hat{U}\sub{shift}(j-i)$.

Next, if we aim to directly measure the matrix element $E_{ij}:=\bracketii{i}{\hat{E}}{j}=\tr(\hat{E}\ket{j}\bra{i})$ of the measurement POVM element $\hat{E}$, the diagrammatic representation suggests that we can select $\hat{A}$, $\hat{\rho}$, and $\hat{B}$ in a manner that $\hat{A}\hat{\rho}\hat{B}^\dag\propto\ket{j}\bra{i}$ holds.
For instance, as demonstrated in Refs.~\cite{PhysRevLett.127.180401,xu2021direct}, if we set $\hat{\rho}=\ket{\mathrm{MUB}}\bra{\mathrm{MUB}}$, $\hat{A}=\ket{j}\bra{j}$, and $\hat{B}=\ket{i}\bra{i}$, the aforementioned condition is satisfied, resulting in $\tr(\hat{A}\hat{\rho}\hat{B}^\dag\hat{E}) = E_{ij}/d$.
As another example, if $\hat{A}$ and $\hat{B}$ are chosen as unitary operators, $\hat{\rho}=\ket{i}\bra{i}$, $\hat{A}=\hat{U}\sub{shift}(j-i)=\sum_k\ket{k+j-i\ \mathrm{mod}\ d}\bra{k}$, and $\hat{B}=\hat{1}$, as shown in Fig.~\ref{fig:2}(b), then the coefficient of $1/d$ vanishes and $\tr(\hat{A}\hat{\rho}\hat{B}^\dag\hat{E}) = E_{ij}$ is obtained. Therefore, the generalized Hadamard test can also yield the expectation value of a measurement POVM element $\hat{E}$ for an unphysical quantum state, which does not adhere to the definition of a density operator such as $\ket{j}\bra{i}$.

Moreover, DM of matrix elements of quantum processes can also be achieved using the generalized Hadamard test. 
The transformation of a quantum process $\mathcal{M}$ for a state $\hat{\rho}$ can be expressed in the following operator-sum representation:
\begin{align}
 \mathcal{M}(\rho)
=\sum_{ijkl}\chi_{ijkl}
\ket{l}\bra{i}\hat{\rho}\ket{j}\bra{k}.
\end{align}
Here we consider DM of the matrix element $\chi_{ijkl}\in\mathbb{C}$. 
$\chi_{ijkl}$ is expressed as follows: 
\begin{align}
 \chi_{ijkl} = \tr\big[\mathcal{M}(\ket{i}\bra{j})\ket{k}\bra{l}\big].
\end{align}
In other words, $\chi_{ijkl}$ can be interpreted as the expectation value of the unphysical measurement POVM element $\ket{k}\bra{l}$ for the unphysical quantum state $\ket{i}\bra{j}$ with time evolution $\mathcal{M}$. 
Therefore, $\chi_{ijkl}$ can be measured using the generalized Hadamard test with gates positioned before and after the quantum process $\mathcal{M}$, as depicted in Fig.~\ref{fig:2}(c). 
Each operator is selected such that $\hat{A}\hat{\rho}\hat{B}^\dag\propto\ket{i}\bra{j}$ and  $\hat{D}^\dag\hat{E}\hat{C}\propto\ket{k}\bra{l}$.
For instance, when using projection operators, we can opt for $\hat{A}=\ket{i}\bra{i}$, $\hat{B}=\ket{j}\bra{j}$, $\hat{C}=\ket{l}\bra{l}$, $\hat{D}=\ket{k}\bra{k}$, and $\hat{\rho}=\hat{E}=\ket{\mathrm{MUB}}\bra{\mathrm{MUB}}$. 
The result of this measurement is given as
\begin{align}
\tr\big[\mathcal{M}(\hat{A}\hat{\rho}\hat{B}^\dag)\hat{D}^\dag\hat{E}\hat{C}\big] = \frac{\chi_{ijkl}}{d^2}.
\end{align}
On the other hand, in the scenario of employing the basis-shift unitary transformation $\hat{U}\sub{shift}$, we can choose $\hat{A}=\hat{U}\sub{shift}(i-j)=\sum_m\ket{m+i-j\ \mathrm{mod}\ d}\bra{m}$, $\hat{B}=\hat{1}$, $\hat{\rho}=\ket{j}\bra{j}$, $\hat{C}=\hat{U}\sub{shift}(k-l)=\sum_m\ket{m+k-l}\bra{m}$, $\hat{D}=\hat{1}$, and $\hat{E}=\ket{k}\bra{k}$. 
Then the coefficient of $1/d^2$ vanishes and 
\begin{align}
\tr\big[\mathcal{M}(\hat{A}\hat{\rho}\hat{B}^\dag)\hat{D}^\dag\hat{E}\hat{C}\big] = \chi_{ijkl}
\end{align}
is obtained. 

\section{Experiment}\label{sec:3}


In this section, we experimentally demonstrate the DM method for matrix elements of quantum states depicted in Fig.~\ref{fig:2}(a). 
This method is one of the DM techniques utilizing the basis-shifted unitary transformation derived through the diagrammatic representation. 
We employ quantum states of photons encoded in discrete time degrees of freedom as the quantum states to be measured. 
In this experiment, we examine three-dimensional quantum states encoded in a series of three laser pulse trains. 
Since the detection probabilities for a single photon, without quantum entanglement, are proportional to the intensity measurement result for classical coherent light, this experiment was conducted using coherent light with output power in the classical regime. 
This experimental approach can be adapted for single photons by employing a suitable single-photon light source and photon detector.


\begin{figure*}[t]		
\centering
\includegraphics[width=15cm]{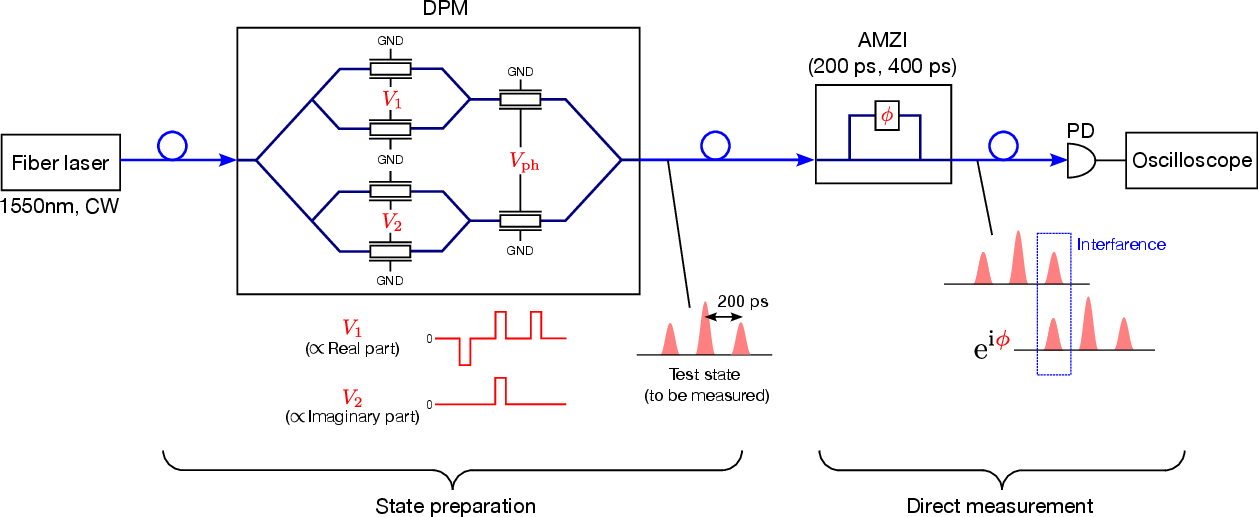}
\caption{
Experimental setup used in the proposed experiment for DM of quantum states. 
Continuous light with a wavelength of 1550\,nm emitted from a fiber laser is injected into a DPM through an optical fiber. 
In the DPM, arbitrary complex amplitudes can be realized by three applied voltages $V_1$, $V_2$, and $V\sub{ph}$. 
Here, we fix $V\sub{ph}=V_{\pi, \mathrm{ph}}$ and apply the voltages corresponding to the real and imaginary parts of the desired complex amplitude to $V_1$ and $V_2$, respectively, to obtain the desired complex amplitude. 
After passing through the DPM, the test pulse enters the DM circuit. 
The waveforms without and with AMZI, which has delays of 200\,ps and 400\,ps, are measured using an oscilloscope. 
The measurements are made for four cases of AMZI phase difference $\phi$: $0^\circ$, $90^\circ$, $180^\circ$, and $270^\circ$.
}
\label{fig:3}
\end{figure*}

\begin{figure}[t]		
\centering
\includegraphics[width=8.5cm]{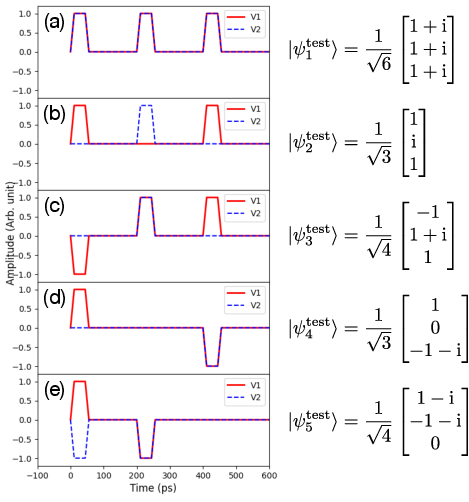}
\caption{
(a)--(e) Waveforms of the voltages $V_1$ and $V_2$ applied to the DPM to realize the test states $\ket{\psi_1\sur{test}}$--$\ket{\psi_5\sur{test}}$.
The pulse repetition period is 200\,ps and the pulse width is 44\,ps.
}
\label{fig:4}
\end{figure}

Figure~\ref{fig:3} illustrates the experimental setup. 
A fiber laser source (NP Photonics SMPF-2030) emits continuous light at a wavelength of 1550\,nm. 
The output laser beam is amplified to 53\,mW by an erbium-doped fiber amplifier (EDFA; GIP Technology), with polarization adjustment performed by a polarization controller (CPC900, Thorlabs; not shown in Fig.~\ref{fig:3}) to optimize injection efficiency into the next component, a dual parallel modulator (DPM; Mach-10 086, COVEGA). 
Subsequently, the DPM modulates the time distribution of the laser light into a triple-pulse test waveform, crucial for verifying the DM method. 
In the DPM, the output complex amplitude is modulated as
\begin{align}
E\sub{out} \propto \sin\left(\pi\frac{V_1}{V_{\pi,1}}\right) 
+ \exp
\left(\ii\pi\frac{V\sub{ph}}{V\sub{\pi,ph}}\right)
\sin\left(\pi\frac{V_2}{V_{\pi,2}}\right)
\end{align}
by applying the three voltages $V_1$, $V_2$, and $V\sub{ph}$ as illustrated in Fig.~\ref{fig:3}, where $V_{\pi, i}\ (i=1,2,\mathrm{ph})$ represents the voltage required for a phase shift of $\pi$ at each voltage application port.
With $V\sub{ph}$ fixed at $ V\sub{\pi, ph}/2$ while assuming $V_1\ll V_{\pi, 1}$ and $V_2\ll V_{\pi, 2}$, we can approximately express:
\begin{align}
E\sub{out} \propto \pi\frac{V_1}{V_{\pi,1}} 
+ \ii\pi\frac{V_2}{V_{\pi,2}}.
\end{align}
Thus, arbitrary complex amplitude waveforms can be generated by applying RF voltages to $V_1$ and $V_2$ that are proportional to the real and imaginary parts, respectively, of the complex amplitude waveform to be generated. 
We have generated five amplitude states from $\ket{\psi\sur{test}_1}$ to $\ket{\psi\sur{test}_5}$, presented in Fig.~\ref{fig:4} as test waveforms. 

The generalized Hadamard test is implemented using an asymmetric Mach–Zehnder interferometer (AMZI) with delays of 200\,ps and 400\,ps. 
AMZI can introduce not only delays between two paths but also the phase difference $\phi$.
We conducted measurements for four phase difference cases: $\phi=0^\circ$, $90^\circ$, $180^\circ$, and $270^\circ$.
From the difference between the measurement results for $\phi=0^\circ$ and $180^\circ$, and those for $\phi=90^\circ$ and $270^\circ$, the expectation values of $\hat{X}$ and $\hat{Y}$ are obtained, respectively. 
The pulse trains output by AMZI were detected by a photodetector (PD), and their time waveforms were observed using an oscilloscope. 

\begin{figure*}[t]		
\centering
\includegraphics[width=17cm]{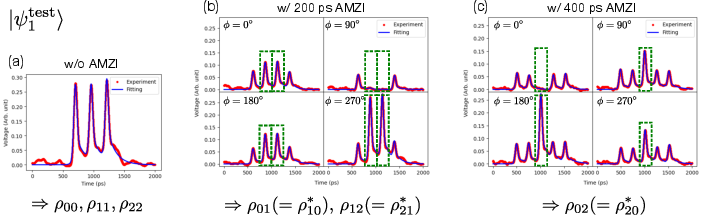}
\caption{
FIG. 5. Experimental procedures for DM of each matrix element within the density matrix of the state $\ket{\psi_1\sur{test}}$. 
In each graph, red dots represent experimental data points, while blue lines depict the fitted curves. 
(a) Measurement outcomes aimed at determining the diagonal components $\rho_{00}$, $\rho_{11}$, and $\rho_{22}$ of the density matrix. 
Without the AMZI, the normalized peak amplitudes directly correspond to $\rho_{00}$, $\rho_{11}$, and $\rho_{22}$. 
(b) Measurement results for obtaining $\rho_{01}$, $\rho_{10}$, $\rho_{12}$, and $\rho_{21}$ within the density matrix. 
In the measurement results for the case of inserting AMZI with 200\,ps delay, the two peak amplitudes enclosed by the green dashed lines are normalized, and the difference between these normalized amplitudes when $\phi=0^\circ$ and $\phi=180^\circ$, $\phi=90^\circ$ and $\phi=270^\circ$ correspond to the real and imaginary parts of the matrix elements to be obtained, respectively. 
The $\rho_{10}$ and $\rho_{21}$ are determined as the complex conjugates of $\rho_{10}$ and $\rho_{12}$, respectively.
(c) Measurement outcomes to determine $\rho_{02}$ and $\rho_{20}$ within the density matrix.
In the measurement results for the case of inserting AMZI with 400\,ps delay, the two peak amplitudes enclosed by the green dashed lines are normalized, and the difference between these normalized amplitudes when $\phi=0^\circ$ and $\phi=180^\circ$, $\phi=90^\circ$ and $\phi=270^\circ$ correspond to the real and imaginary parts of $\rho_{02}$, respectively. 
$\rho_{20}$ is deduced as the complex conjugate of $\rho_{02}$.
}
\label{fig:5}
\end{figure*}

Taking the case of $\ket{\psi_1\sur{test}}$ as an example, we explain the procedure for obtaining each matrix element from the observed waveform in this DM method. 
Initially, we examine the waveform without AMZI, as depicted in Fig.~\ref{fig:5}(a), and derive three pulse amplitudes through fitting, denoted as $A_0$, $A_1$, and $A_2$, respectively (refer to Appendix~\ref{appendix:A} for details regarding the fitting function). 
Normalizing these values allows us to obtain the diagonal components of the matrix as follows:
\begin{gather}
A:=A_0 + A_1 + A_2\\
\rho_{00} = \frac{A_0}{A},\quad
\rho_{11} = \frac{A_1}{A},\quad
\rho_{22} = \frac{A_2}{A}. 
\end{gather}

Next, as depicted in Fig.~\ref{fig:5}(b), the waveform shifted by one pulse using the basis-shift unitary transformation is interfered with the original waveform by employing AMZI with a delay of 200\,ps. 
We observe the waveform after passing through AMZI and obtain the amplitudes of the second and third pulses by fitting. 
Let $A_{01}(\phi)$ and $A_{12}(\phi)$ represent the amplitudes of the second and third pulses at the phase difference $\phi$.
%
Using the normalization constant $A$ obtained above, we obtain the values of $\rho_{01}$ and $\rho_{12}$ as follows:
\begin{align}
 \rho_{01} &= \frac{[A_{01}(0^\circ)-A_{01}(180^\circ)] + \ii\left[A_{01}(90^\circ) - A_{01}(270^\circ)\right]}{A},\\
\rho_{12} &= \frac{[A_{12}(0^\circ)-A_{12}(180^\circ)] + \ii\left[A_{12}(90^\circ) - A_{12}(270^\circ)\right]}{A}.
\end{align}
The values of $\rho_{10}$ and $\rho_{21}$ are obtained as complex conjugates of $\rho_{01}$ and $\rho_{12}$, respectively. 

\begin{figure*}[t]		
\centering
\includegraphics[width=17cm]{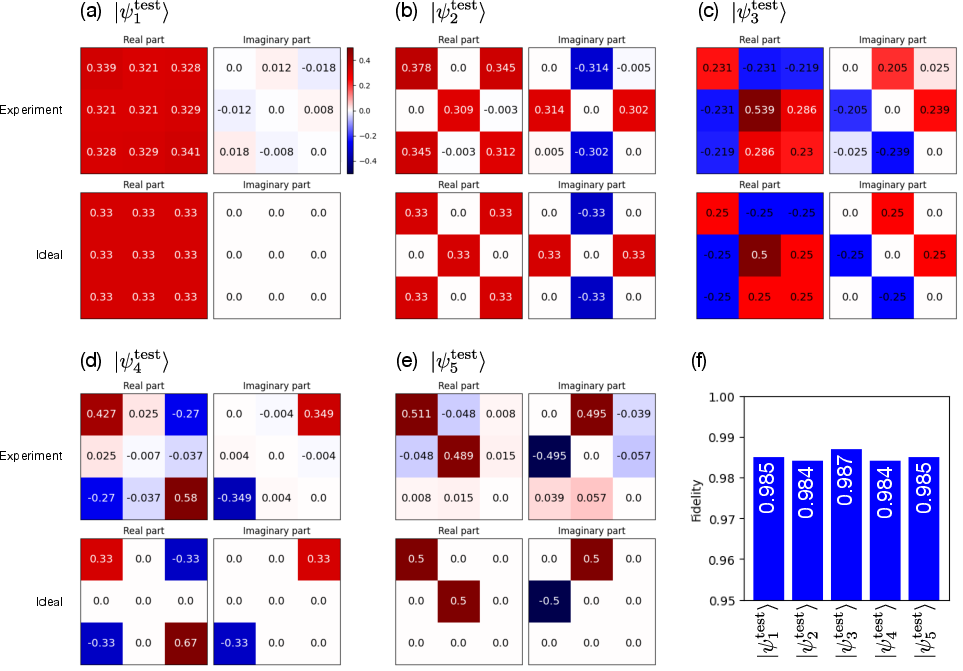}
\caption{
(a)--(e) Results of DM of the density matrix for the quantum state $\ket{\psi_1\sur{test}}$--$\ket{\psi_5\sur{test}}$ (top panel) and ideal values (bottom panel). (f) Fidelities of the non-negative and normalized density matrices obtained by DM to the ideal states. 
}
\label{fig:6}
\end{figure*}

Subsequently, as depicted in Fig.~\ref{fig:5}(c), we observe the waveforms after passing through the AMZI with a delay of 400\,ps, equivalent to a basis shift of two pulses, and extract the amplitude of the third pulse through fitting. 
Denoting $A_{02}(\phi)$ as the amplitude of the third pulse at the phase difference $\phi$, $\rho_{02}$ is obtained as follows:
\begin{align}
 \rho_{02} &= \frac{[A_{02}(0^\circ)-A_{02}(180^\circ)] + \ii\left[A_{02}(90^\circ) - A_{02}(270^\circ)\right]}{A}.
\end{align}
Similarly, $\rho_{20}$ is obtained as the complex conjugate of $\rho_{02}$. 
This procedure allows us to derive every matrix element of the density operator of the quantum state. 
Figs.~\ref{fig:6}(a)--(e) present the results of DM of the density matrix for each state.
Notably, each matrix element obtained through the direct measurement method closely approximates the ideal value, demonstrating the efficacy of the proposed DM method.

Furthermore, the density matrix $\hat{\rho}\sur{exp}$ reconstructed by this DM method is evaluated by the fidelity:
\begin{align}
F(\hat{\rho}\sur{exp}, \ket{\psi\sur{test}}) = \sqrt{\bracketii{\psi\sur{test}}{\hat{\rho}\sur{exp}}{\psi\sur{test}}},
\end{align}
where $\ket{\psi\sur{test}}$ represents the ideal state. 
The matrices presented in Figs.~\ref{fig:6}(a)--(e), obtained through the current method, do not adhere to the requirement of non-negativity and unity trace that a density operator should satisfy.
Consequently, we applied a transformation to ensure that the matrices obtained via the DM method conform to the properties of non-negativity and unity trace (refer to Appendix~\ref{appendix:B} for details regarding the transformation procedure and the resulting matrices). 
The fidelities of these modified matrices with respect to the ideal states are depicted in Fig.~\ref{fig:6}(f).  
Notably, all fidelities exceed 0.98, indicating the proper functionality of the proposed direct measurement method.

\section{Discussion and summary}\label{sec:4}
 

The DM method utilizing the basis-shift unitary transformation, proposed in this paper, proves to be the most efficient in terms of DM of matrix elements using qubit probes. 
It minimizes the statistical fluctuation of the estimated value to be measured within a fixed number of shots. 
This efficiency can be explained as follows: In measurements with qubit probes, the value to be estimated, denoted as $A\in[-1,+1]$, is given in proportion to the expectation value of the Pauli-$Z$ measurement:
\begin{align}
A = k\bracket{\hat{Z}} = k(p_+ - p_-) = k(2p_+-1),\label{eq:7}
\end{align}
where $p_\pm$ is the probability that the Pauli-$Z$ measurement yields the outcome $\pm 1$. 
The proposed DM method with basis-shift unitary transformation corresponds to the case $k=1$.
Let $\bar{Z}_n$ be the sample mean of $Z$ measurements taken $n$ times.  
The estimator $\tilde{A}_n$ of $A$ is given by $\tilde{A}_n = k\bar{Z}_n$. 
Using Eq.~(\ref{eq:7}), the sample variance of $\bar{Z}_n$ is given as
\begin{align}
 \mathrm{Var}(\bar{Z}_n)
=\frac{\mathrm{Var}(\hat{Z})}{n}
=\frac{4p_+(1-p_+)}{n}
=\frac{1-k^{-2}A^2}{n},
\end{align}
and the sample variance of $\tilde{A}_n$ is expressed as 
\begin{align}
\mathrm{Var}(\tilde{A}_n)
=k^2\mathrm{Var}(\bar{Z}_n)
=\frac{k^2-A^2}{n}.
\end{align}

Given that the variance is minimized when $k=1$, it is concluded that the proposed DM method employing the basis-shift unitary transformation stands as the most efficient approach for DM of matrix elements using qubit probes. 
In contrast, in most previous studies \cite{lundeen2011direct,PhysRevLett.112.070405,PhysRevLett.117.120401,PhysRevLett.123.150402,kim2018direct,gaikwad2023direct,PhysRevLett.127.180401,PhysRevLett.116.040502,PhysRevLett.121.230501,ogawa2021direct,PhysRevLett.127.040402,PhysRevA.101.012119,PhysRevLett.127.030402,xu2021direct}, $k<1$ since they conduct the projection measurement in MUB for the basis of the matrix to be obtained. 
Hence, the efficiency of the proposed DM method represents a notable advantage. 
The use of basis-shift unitary transformation is also expected to improve the measurement efficiency in DM of processes and measurements. 


In conclusion, we have introduced a theoretical framework aimed at unifying the DM methods for the three components of quantum dynamics: quantum states, quantum processes, and quantum measurements. 
Leveraging this framework, DM methods for each quantum component can be systematically derived. 
Particularly, we have developed a DM method utilizing the basis-shift unitary transformation, which stands out as the most efficient approach when utilizing qubit probes. 
Furthermore, we have conducted experimental demonstrations of the DM method for quantum states using optical pulse trains. 
Our experiments have showcased that the matrix elements of the density operator of the initial state can be obtained separately, and the overall state can be estimated with high fidelity. 
Although this experiment utilized optical pulse trains for quantum communication applications, we anticipate that this DM method will find applications in various physical systems. 
We envisage that in the future, DM techniques will not only facilitate the measurement of states but also enable the measurement of processes and measurements in diverse quantum systems.


\begin{acknowledgments}
This research was supported by JSPS KAKENHI Grant Number 21K04915.
\end{acknowledgments}

\clearpage

\appendix

\begin{widetext}

\section{Details of experimental results and fitting process}\label{appendix:A}


In the experiments detailed in Section~\ref{sec:3}, we produced optical pulse trains corresponding to the five initial states $\ket{\psi_1^{\rm test}}$--$\ket{\psi_5^{\rm test}}$ and detected these pulses using an oscilloscope following the proposed DM procedure. 
By fitting a function with a peak shape to the detected waveforms, we obtained the values of the peaks of each pulse. 
The measured waveforms and the fitting results are depicted in Fig.~\ref{fig:7}. 
For each individual peak, the following function is employed as a peak shape fitting function:
\begin{align}
f_{\rm fitting}(x) = A\left\{
\exp\left[
\frac{-(x-x_0)^2}{2\sigma^2}
\right]
+ R
\frac{\exp[-(x-x_0)/\tau_1]}{1+\exp[-(x-x_0)/\tau_2]}
\right\},
\end{align}
where the first term represents a Gaussian function with a center at $x_0$ and a standard deviation of $\sigma$, while the second term represents a peak shape function with a center at $x_0$ and exhibits asymmetric behavior with its left and right sides decreasing exponentially with decay constants $\tau_1$ and $\tau_2^{-1}-\tau_1^{-1}$, respectively.  
The magnitude ratio of the second term to the first term is denoted by $R$. 
The second term is added to express the asymmetric shape of the measured results for each peak, with a slower decay on the right side. 
The fitting functions were obtained by adding three, four, and five instances of the aforementioned peak functions for cases without the AMZI, with AMZI inserted with a delay of 200\,ps, and with AMZI inserted with a delay of 400\,ps, respectively. 
Each peak was fitted to the measured data to obtain parameters including $x_0, \sigma, \tau_1, \tau_2, A$, and $R$. 
These peak values $A$ were then utilized to estimate each matrix element of the initial state according to the procedure outlined in the main text.

\begin{figure*}[t]		
\centering
\includegraphics[width=15cm]{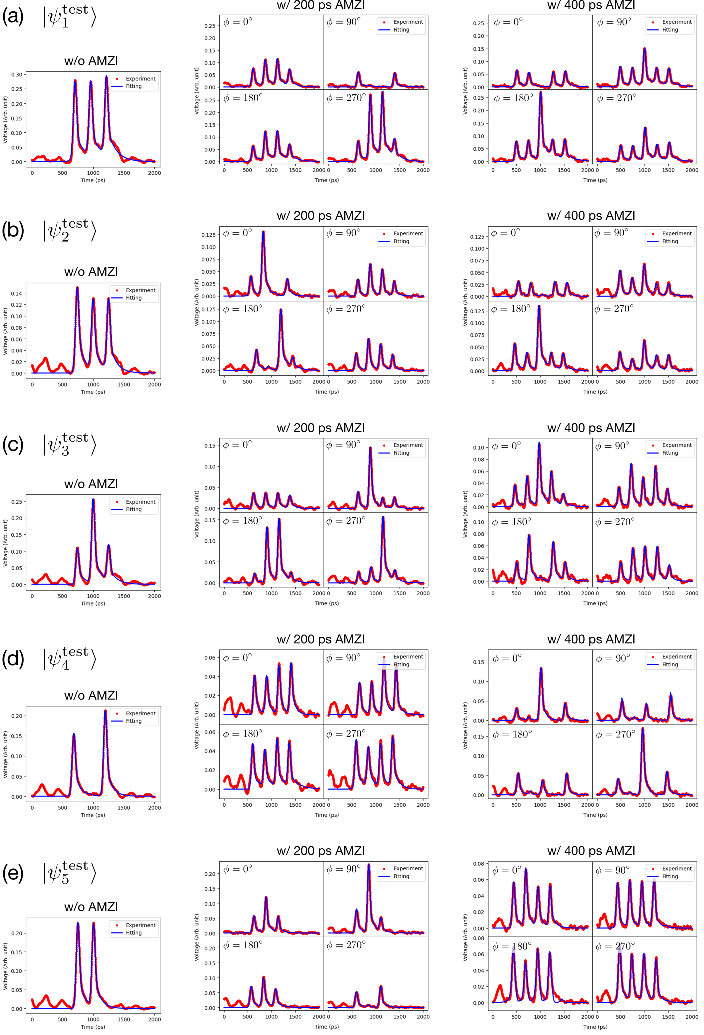}
\caption{
(a)--(e) Measured waveforms and fitting results for the five initial states $\ket{\psi_1^{\rm test}}$--$\ket{\psi_5^{\rm test}}$. 
In each plot, red dots represent experimental measurements, and blue lines depict fitting curves. 
Left panels: measurement results without the AMZI. 
Center panels: measurement results with AMZI inserted with a 200\,ps delay and phase differences of of $\phi=0^\circ$, $90^\circ$, $180^\circ$, and $270^\circ$. 
Right panels: measurement results with AMZI inserted with a 400\,ps delay and phase differences of $\phi=0^\circ$, $90^\circ$, $180^\circ$, and $270^\circ$.
}
\label{fig:7}
\end{figure*}

\section{Post-processing of density matrices measured by direct measurement}\label{appendix:B}


When each matrix element is obtained individually by DM, the resultant matrices generally fail to meet the physical requirements of density operators, namely being non-negative operators with a unity trace. 
To rectify this in the experiment, we post-processed the density matrix $A$ obtained by DM using the following steps: 
First, to ensure the non-negativity requirement, $A$ underwent spectral decomposition, and the imaginary part of its eigenvalues was replaced by zero.
If the real part was negative, the real part was also replaced by zero. 
Subsequently, to ensure the unit trace requirement, the entire matrix was normalized to $A\rightarrow A/a$, where $a$ is the sum of the diagonal components $a:=A_{00} + A_{11} + A_{22}$. 
The matrices obtained after this post-processing are displayed in Fig.~\ref{fig:8}. 
The fidelities of these matrices with respect to the ideal states were evaluated as discussed in the main text.

\begin{figure*}[t]		
\centering
\includegraphics[width=17cm]{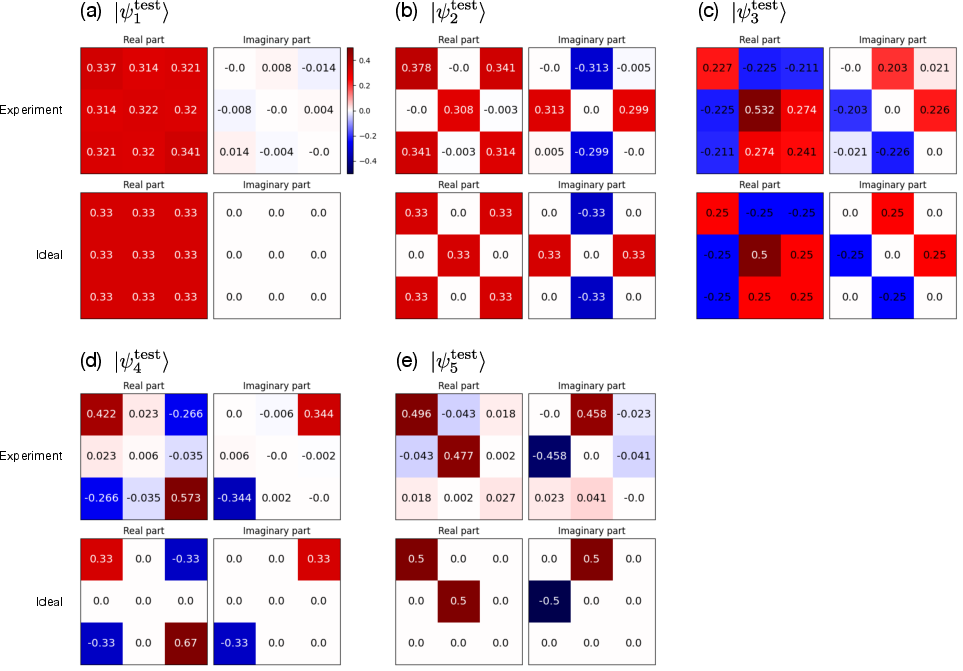}
\caption{
(a)--(e) Density matrices of the quantum state $\ket{\psi_1\sur{test}}$--$\ket{\psi_5\sur{test}}$ measured by DM, post-processed to meet the requirements of a non-negative operator and unit trace (upper panel), alongside the ideal values (bottom panel).
}\label{fig:8}
\end{figure*}

\end{widetext}

\end{spacing}

\nocite{*}
\bibliography{ref}

\end{document}